\documentclass[%
 reprint,
 amsmath,amssymb,
 aps,
 prl
]{revtex4-2}

\usepackage{graphicx}
\usepackage{dcolumn}
\usepackage{bm}
\usepackage[dvipsnames]{xcolor}
\usepackage{siunitx}
\usepackage{hyperref}

\begin{document}

\preprint{APS/123-QED}

\title{A Fermi-Hubbard Optical Tweezer Array}

\author{Benjamin M. Spar}
\affiliation{Department of Physics, Princeton University, Princeton, New Jersey 08544, USA}
\author{Elmer Guardado-Sanchez}
\altaffiliation[Present address: ]{Department of Physics, Harvard University, Cambridge, MA 02138, USA}
\affiliation{Department of Physics, Princeton University, Princeton, New Jersey 08544, USA}
\author{Sungjae Chi}
\affiliation{Department of Physics, Princeton University, Princeton, New Jersey 08544, USA}
\author{Zoe Z. Yan}
\affiliation{Department of Physics, Princeton University, Princeton, New Jersey 08544, USA}
\author{Waseem S. Bakr}
\affiliation{Department of Physics, Princeton University, Princeton, New Jersey 08544, USA}

\date{\today}

\begin{abstract}
We use lithium-6 atoms in an optical tweezer array to realize an eight-site Fermi-Hubbard chain near half filling. We achieve single site detection by combining the tweezer array with a quantum gas microscope. By reducing disorder in the energy offsets to less than the tunneling energy, we observe Mott insulators with strong antiferromagnetic correlations. The measured spin correlations allow us to put an upper bound on the entropy of 0.26(4)\,$k_\mathrm{B}$ per atom, comparable to the lowest entropies achieved with optical lattices.  Additionally, we establish the flexibility of the tweezer platform by initializing atoms on one tweezer and observing tunneling dynamics across the array for different 1D  geometries.
\end{abstract}

\maketitle

\begin{figure*}[t]
\includegraphics[width=\textwidth]{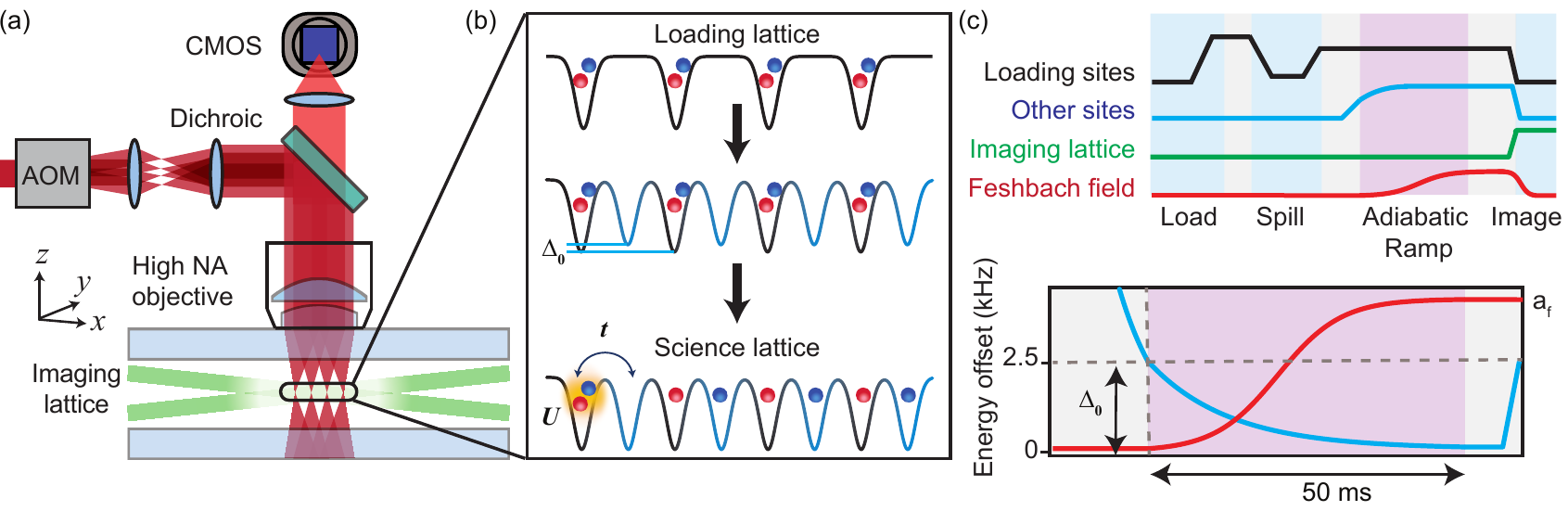}
\caption{\label{fig:1}Experimental setup for the Fermi-Hubbard tweezer array. (a) Schematic diagram of the experiment. The tweezer array is generated by an AOM and is projected through an objective. Correlated states of the Hubbard model are prepared in the tweezer array and the atoms are subsequently loaded from the tweezers into a two dimensional optical lattice in the $x-y$ plane for imaging. A lightsheet potential provides vertical confinement along $z$ during imaging. Light scattered during Raman cooling in the 2D lattice is captured through the same objective.  (b) Protocol for creating low entropy correlated states. First, we load the ground state of well-separated tweezers with two atoms in different spin states using a spilling procedure with a magnetic gradient. Next, additional tweezers ramp on to an energy offset $\Delta_0$ where there is no tunneling. Finally, the energy offset between neighboring tweezers is reduced to zero to adiabatically prepare an antiferromagnetic state.  (c) Experimental sequence. To initialize the Fermi-Hubbard array, the energy offset between neighboring tweezers is decreased to zero while the scattering length is increased to its final value $a_f$, reaching the desired $U/t$.}
\end{figure*}

Arrays of neutral atoms in optical tweezers have emerged as a powerful new platform for quantum simulation and computation \cite{barredo2016atom,endres2016atom}. It is now possible to study interacting quantum systems in one and two-dimensional tweezer arrays, largely defect-free, with hundreds of atoms \cite{scholl2020programmable,ebadi2020quantum}. Initial work with reconfigurable tweezer arrays used alkali atoms, and the platform has now expanded to include alkaline earth atoms \cite{cooper2018alkaline,norcia2018microscopic,saskin2019narrow} and molecules \cite{liu2018building,anderegg2019optical}. Notable results include studies of quantum spin models using atoms excited to Rydberg states \cite{labuhn2016tunable, bernien2017probing}, demonstration of high fidelity quantum gates \cite{levine2019parallel, madjarov2020high}, and high quality factor atomic clocks \cite{young2020half}. Most of the activity with tweezers arrays has focused on atoms localized on individual tweezers. The versatility of tweezer arrays provides a strong incentive for extending quantum simulations with this platform to systems of mobile atoms where the effects of quantum statistics become important. A key step in this direction has been the demonstration of tunnel-coupled double-well tweezer systems \cite{kaufman2014two,murmann2015two}. 

The use of tweezer arrays to study itinerant condensed matter models such as the Hubbard model realizes a ``bottom-up" paradigm of quantum simulation, in contrast to the more established ``top-down" approach of using optical lattices \cite{bloch2005ultracold}. Optical lattices are an efficient way to create periodic trapping potentials with thousands of lattices sites that can be loaded directly from a degenerate gas. In recent years, quantum gas microscopes have been used to probe optical lattices with single-site resolution, allowing for extraction of multi-point correlation functions in real space. In particular, fermionic quantum gas microscopes have been used to explore the phase diagram of the square lattice Fermi-Hubbard model, a minimal model for high temperature superconductivity \cite{taruell2018}. Quantum gas microscopes have allowed for direct measurement of the Mott insulator state \cite{greif2016site, CheukMott}, antiferromagnetic correlations at half filling \cite{cheuk2016observation,parsons2016site,boll2016spin, brown2017spin}, and the motion of a single hole in an antiferromagnetic background \cite{vijayan2020time,ji2021dynamical,koepsell2019Imaging}. 

Studying Fermi-Hubbard models with optical lattices faces two challenges that motivate the consideration of tweezer arrays as an alternative platform. First, the lowest entropies that have been achieved for correlated states in optical lattices are in the range of $0.3{-}0.5\,k_\mathrm{B}$ per particle~\cite{mazurenko2017cold,brown2019bad,sompet2021realising}. This has hindered access to interesting regimes of the square Hubbard model phase diagram such as the pseudogap or the putative $d$-wave superconductor \cite{RevModPhys.78.17}. Sophisticated entropy redistribution schemes have been investigated, but were limited by the ability to precisely control lattice potentials at the single-site level~\cite{chiu2018quantum}. A second challenge, particularly relevant for microscope experiments, is the difficulty of reconfiguring the apparatus to study different lattice geometries. Programmable Fermi-Hubbard tweezer arrays have the potential to address both of these issues by allowing precise dynamical control of the simulated model at the single site level. This includes the geometry of the array, energy offsets on individual tweezers, and the tunneling matrix element on each bond. Going beyond square Hubbard models will enable microscopic studies of qualitatively different phenomena in correlated systems including flat bands, Dirac points, and quantum spin liquids. 

The requirements for observing coherent tunneling between two tweezers is the ability to prepare atoms in the motional ground state and to control the energy offset between the tweezers to better than the tunneling energy, which is normally less than a percent of the total depth. In Ref.~\cite{murmann2015two}, a two-site Fermi-Hubbard model was realized by loading a pair of tweezers with atoms from a degenerate Fermi gas. These experiments have been limited in expanding to large arrays due to the difficulty of \textit{in-situ} imaging of $^6$Li in optical tweezers \cite{bergschneider2018spinresolved}. In this work, we combine a tweezer array with a quantum gas microscope to study a programmable eight-site Hubbard chain, an increase in the size of the Hilbert space by over three orders of magnitude.  

We implement the Fermi-Hubbard model with two hyperfine spin states of $^6$Li loaded in a one-dimensional tweezer array. By loading the ground state of four independent tweezers with high fidelity, we adiabatically transform a low-entropy band insulator into a correlated state by ramping on four additional tunnel-coupled tweezers to change the filling of the system. This scheme is similar to proposed adiabatic preparation using an optical superlattice that allows for single site control of the ramping procedure \cite{Lubasch}. Near the end of the ramp, the system is well-described by the single-band Hamiltonian
\begin{align}
    \hat{H}_{FH} = &- \sum_{\langle i, j\rangle, \sigma} t_{ij} (\hat{c}^{\dag}_{i \sigma} \hat{c}_{j \sigma} + h.c.) \nonumber\\
    &+ U \sum_{i} \hat{n}_{i \uparrow} \hat{n}_{i \downarrow} + \sum_{i, \sigma} \Delta_i \hat{n}_{i \sigma},
\end{align}
where $\hat{c}^{\dag}_{i \sigma}$ is the fermionic creation operator of spin $\sigma$ at site $i$, and $\hat{n}_{i\sigma}$ is the number operator.  Here, the local tunneling matrix element $t_{ij}$, energy offset  $\Delta_i$, and on-site interaction $U$ can be controlled in real time. By carefully controlling the Hubbard parameters with tweezer spacing, depth, and the interparticle scattering length, we prepare low entropy states with antiferromagnetic correlations, showcasing the ability of the tweezer platform to generate ``clean" many-body systems.

We generate the tweezer array with 770 nm light using an acousto-optical modulator (AOM), such that different tweezers are generated by radio-frequency tones of different frequencies (Fig.~\ref{fig:1}(a)). We implement elliptic tweezers with a waist of $\approx\,$930 nm at the atoms as measured along the direction of the tweezer array and a waist of $\approx\,$1250 nm in the perpendicular direction. We work with two different configurations of the tweezers that we switch between during an experimental cycle: the loading configuration with independent tweezers and the science configuration with tunnel-coupled tweezers at half the separation of the loading configuration. In the science configuration, adjacent tweezers differ in radio-frequency tone by 4 MHz, corresponding to a lattice spacing of 1350 nm. The beat frequency of neighboring tweezers is about two orders of magnitude larger than the tweezer depths, leading to negligible parameteric heating. The aperture size and the bandwidth of the AOM limit us to a tweezer array with eight sites. We load the initial configuration of four tweezers from an attractively interacting degenerate gas that is an equal mixture of the lowest and third lowest hyperfine ground states of $^6$Li prepared in an optical dipole trap. The initial temperature of the gas is $\approx$ 0.2 times the Fermi temperature, which does not limit the final entropy of the tweezer array \cite{serwane2011deterministic}. After allowing the system to equilibrate, the optical dipole trap is slowly turned off and the magnetic field is ramped to a non-interacting value.

Initially, there are tens of atoms of each spin state occupying the lowest energy levels of each trap. To remove atoms in higher energy levels, we apply a magnetic gradient while lowering the depth of each trap to spill out all atoms except for one atom in each spin state in the ground state, similar to Ref.~\cite{serwane2011deterministic}. Accounting for imaging fidelity, each spin state is loaded with a fidelity $\langle  n_{\uparrow} \rangle = \langle  n_{\downarrow} \rangle = 0.975(9)$~\cite{Supplement}. We bias the spilling procedure such that almost all of the errors in preparation result in one atom per tweezer and we avoid preparing any atoms in excited motional states. After spilling, we quickly ramp on the additional tweezers needed for the science configuration to $\sim 95\%$ of the depth of the loading tweezers, corresponding to an energy offset $\Delta_0$ in Fig.~\ref{fig:1}(b). In the last stage of the experimental sequence, we slowly decrease the tweezer energy offsets to zero in 50\,ms as the scattering length of the atoms is ramped to its final value by increasing the magnetic field (see Fig.~\ref{fig:1}(b),(c)). Before imaging the resulting correlated state, tunneling is frozen in the array by offsetting the tweezers back to $\Delta_0$ in 100$\,\mu$s and increasing the overall depth by a factor of 3. As the eight site Fermi-Hubbard system can be exactly diagonalized, we use the Python package QuSpin to simulate the ramping procedure for comparison to the experiment \cite{weinberg2019quspin}.

Due to its light mass, lithium is a challenging species to image in optical tweezers. Therefore, to detect the atoms with single-site resolution, we transfer them from the tweezer array into a two-dimensional square optical lattice. The lattice has a $752\,$nm spacing and its axes are rotated by $45^{\circ}$ relative to the tweezer array, which allows us to super-resolve the array. At a depth of 2500 $E_R$, the lattice has much larger radial trapping frequencies than the tweezers, which allows us to reach the Lamb-Dicke regime for effective Raman sideband cooling during fluorescence imaging~\cite{brown2017spin}. Using this scheme, we detect atoms with 98.5(4)\% fidelity. 

For typical tunneling energies of $h{\times}200\,$Hz and tweezer depths of $h{\times}50\,$kHz, where $h$ is Planck's constant, we need to balance the intensities of the tweezers to better than 0.5\% of the depth to be in the regime where disorder is comparable to or less than the tunneling. This requirement is more stringent than for experiments with Rydberg atoms, which typically only need to equalize the tweezer depths to within a few percent \cite{endres2016atom, young2020half}. We start by balancing tweezer intensities on a camera, but this is not precise enough because of subsequent intensity differentials introduced as the light passes through the tweezer projection optics.  Therefore, fine balancing of tweezer depths in the science configuration is achieved by feeding back on the the average density profile of the atoms. Working at weak interactions of $U/t \sim 1$ gives  good sensitivity to disorder and acceptable many-body gaps for adiabatic preparation. We measure the average density for a given set of tweezer depths using $40$ experimental runs and then use a proportional feedback loop on the radio-frequency amplitude of each tone based on the associated density distribution to homogenize the science configuration~\cite{Supplement}. For rapid convergence of the feedback, cross couplings between different tones are minimized by operating the AOM and its driver well below saturation. Starting from a typical disorder configuration, this scheme takes thirty minutes to an hour to equalize the tweezer offsets to within half of the tunneling energy, and stays balanced on the day scale. We use a similar scheme to balance tweezer depths in the loading configuration; however, it is a less sensitive procedure since the relevant energy scale for spilling is the axial trapping frequency of the tweezers (${\sim}\,2$ kHz), which is an order of magnitude larger than the tunneling. 

In the offset-balanced science configuration array, the atoms realize the Fermi-Hubbard model with disorder less than the tunneling. As the loading fidelity of the band insulator in the loading configuration is not perfect, the system is on average slightly below half filling. After the ramp to the science configuration, the highest $\langle n \rangle\,{\equiv}\,p$ is 0.955(7), indicating some atom loss during the ramp.

\begin{figure}[t]
\includegraphics{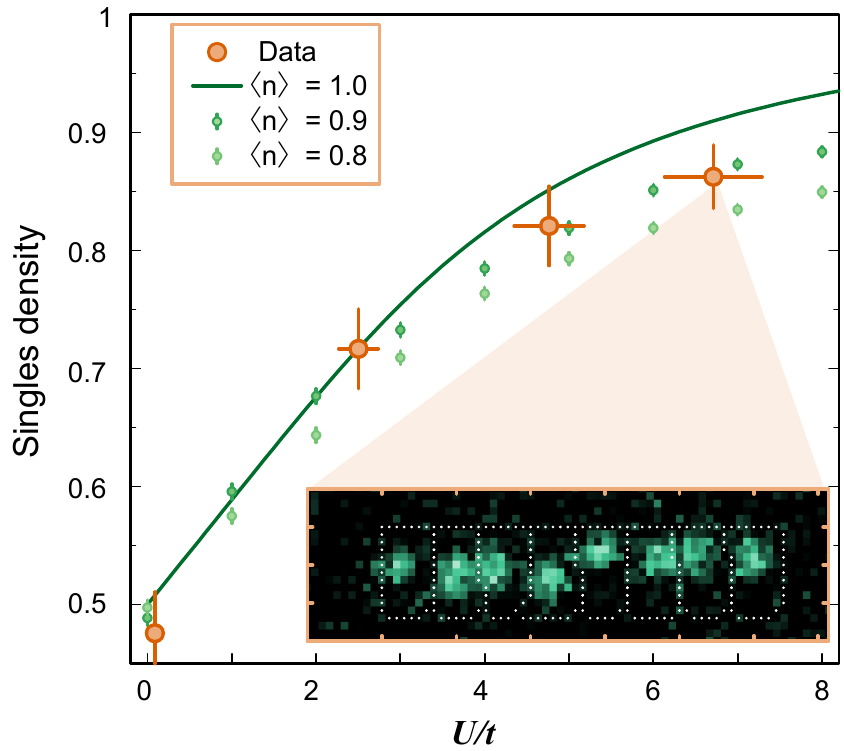}
\caption{\label{fig:2}Singles density as a function of $U/t$. We compare the experimental data to the calculated singles density at three different values of the total density. The measured density $\langle n \rangle\,{=}\, 0.93(2)$ of this dataset is caused by imperfect loading. The inset is a fluorescence image of an eight atom Mott insulator with reconstruction masks (shown as white dotted lines) to identify which tweezer the atom originated from.}
\end{figure}

To verify that we can prepare correlated states in the science configuration, we look for Mott insulators at large repulsive interactions, where it is energetically favorable for atoms to be on singly occupied sites. To image singly occupied sites, we first convert atoms on doubly occupied tweezers into Feshbach molecules to ensure loss before loading into the optical lattice~\cite{Supplement}. We measure the fraction of singly occupied sites in the Mott insulator as a function of interaction energy (Fig.~\ref{fig:2}). For this measurement, $t/h\,{=}\,$160(4) Hz, and the scattering length is tuned from 0 to 1600 Bohr radii ($a_0$). At a scattering length of $1600\,a_0$, we measure $U/h\,{=}\,1.07(4)$ kHz, giving a maximum $U/t$ of $6.7(3)$. Using exact diagonalization methods, we extract the expected singles occupation for different densities, and find that the singles occupation we measure at different interactions is consistent with the density $ \langle  n \rangle \,{=}\,0.93(2)$ of this dataset.  

\begin{figure}[b]
\includegraphics{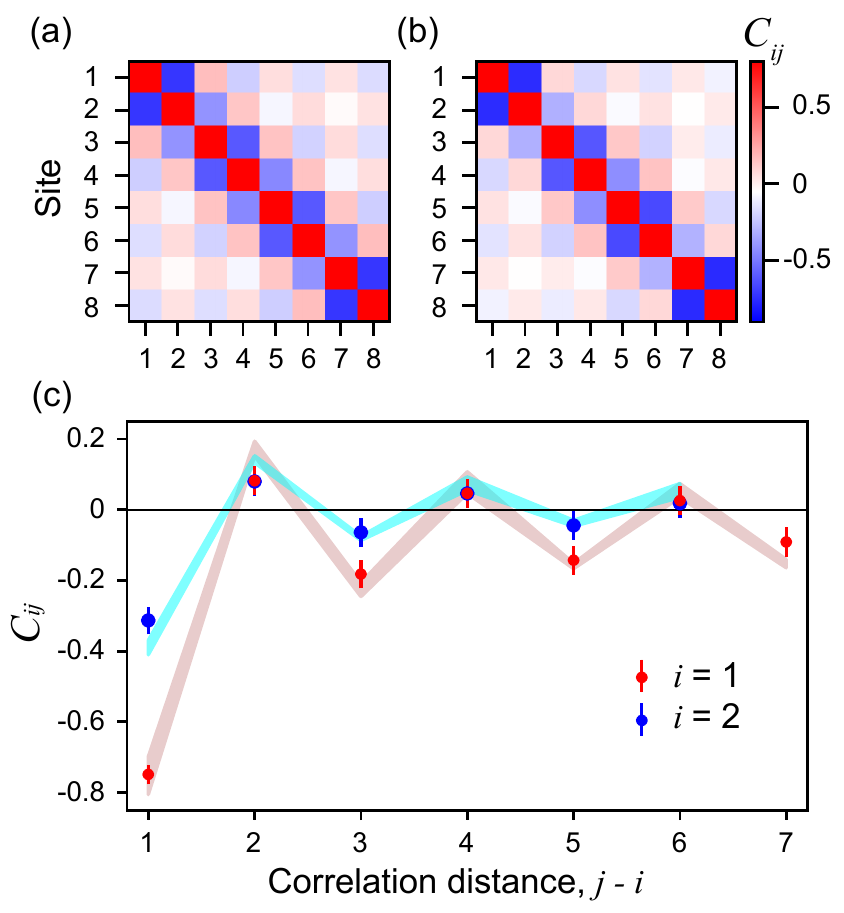}%
\caption{\label{fig:3}Antiferromagnetic correlations in the tweezer array. (a) Up-up antiferromagnetic correlations between all lattice sites calculated with exact diagonalization at a temperature $k_\mathrm{B}T\,{=}\,0.21(3)\,t$ and (b) experimentally measured correlations. Due to finite size effects, the edge-nearest neighbor pairs have stronger correlations than other nearest-neighbor pairs. (c) Correlations as a function of distance for the first site (red) and second site (blue) of the chain, corresponding to the first and second rows in the top plots. The shaded curve is the range of calculated correlation values for a specific pair of sites given by the fitted temperature range.}
\end{figure}

Although the suppression of doublons conclusively demonstrates the formation of a correlated state, measurements of the density alone are insufficient to characterize the system at low temperatures. At these temperatures, the atoms preferentially arrange themselves in an antiferromagnetic configuration because of the superexchange interaction. By removing atoms in one spin state with resonant light and imaging the other state (the lowest hyperfine state, $\left|\uparrow\right\rangle$), we measure up-up density correlations between sites $i$ and $j$ as $C_{ij}\,{=}\,4 ( \langle n_{ i \uparrow} n_{ j \uparrow} \rangle - \langle n_{ i \uparrow} \rangle \langle  n_{ j \uparrow} \rangle )$ (Fig.~\ref{fig:3}). Due to strong quantum fluctuations, spin correlations in the one-dimensional Fermi-Hubbard model decay over a few sites even in the ground state. Furthermore, in our finite size system, the correlations at a fixed distance are not the same for each pair of atoms. For analysis of the correlations, we post-select on imaging four $\left|\uparrow\right\rangle$ atoms, which lowers the effective temperature and increases the filling~\cite{Supplement}. We simulate the system using the grand canonical ensemble, defined by chemical potential $\mu$ and temperature $T$, as the atom number in our experiment fluctuates due to imperfect loading \cite{kaufman2016quantum}. Using a least squares fit on the average atom number and each individual spin correlation, we find a local minimum of temperature at $k_\mathrm{B} T\,{=}\,0.21 (3)t$, where the errorbar is extracted using bootstrapping methods. However, many of the correlators do not have a strong dependence on temperature, and some are non-monotonic with temperature~\cite{Supplement}. Around 10\% of bootstrapped samples fit to a temperature much closer to zero 
than the rest of the samples. These temperatures are even lower than the temperature expected from numerically evolving the initial loading configuration. 
This suggests that spin correlations cease to be a good thermometer at the lowest temperatures we achieve and the fitted temperature is an upper bound for the temperature of our system.

A more natural quantity to discuss when characterizing closed cold atom systems is the entropy. In the loading configuration, after post-selecting on detecting four spin up atoms, the initial entropy is $0.09(1)\,k_\mathrm{B}$ per particle, calculated from
\begin{equation}
 \frac{S}{\langle N\rangle} = -\frac{k_\mathrm{B}}{1+p} \big(p\log p+(1-p) \log (1-p) \big).
\end{equation}
To extract the final entropy after the ramp, we compute the entropy from the fitted grand canonical parameters, obtaining an upper bound on the entropy per particle of $0.26(4)\,k_\mathrm{B}$. This entropy is comparable to the lowest entropies measured in optical lattices with fermionic quantum gas microscopes~\cite{mazurenko2017cold,brown2019bad,sompet2021realising}. To understand where the entropy gain in the system is coming from, we simulate the ramping procedure and find an expected final entropy of $0.18(2)\,k_\mathrm{B}$ per particle, which is a lower bound on the entropy of the system~\cite{Supplement}. By implementing full spin and density readout in future experiments~\cite{koepsell2020robust}, the entropy of the initial state can be eliminated via post-selection. In that case, numerics indicate that non-adiabticity during the ramp would limit the entropy to $0.04\,k_\mathrm{B}$ per particle.

\begin{figure}
\includegraphics{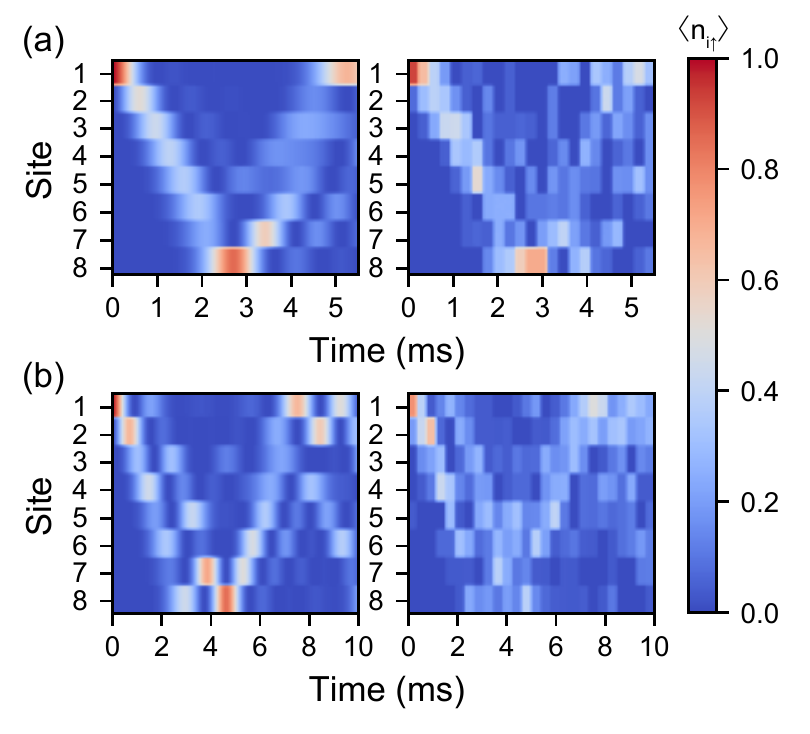}
\caption{\label{fig:4}Pair of non-interacting atoms initialized on one edge site tunneling across the array. (a) Exact diagonalization (left) and experimental (right) dynamics in a configuration with equal tunnelings across all sites. We extract a tunneling rate of 296(3) Hz using a least squares fit. (b) Exact diagonalization (left) and experimental (right) dynamics dynamics with staggered tunneling rates. For site $i$ and site $j= i+1$, $t_{ij} = \SI{289(3)}{\Hz}$ for odd $i$ and $t_{ij} = \SI{213(2)}{\Hz}$ for even $i$. }
\end{figure}

Not only can the tweezer platform prepare highly-correlated low-entropy equilibrium states, but it can also realize dynamics that are difficult to study in comparable optical lattice systems. To demonstrate the flexibility of the platform, we perform experiments where we prepare atoms only on an edge site of the science configuration and observe the propagation of the particles in the array (Fig.~\ref{fig:4}(a)). To initialize dynamics, we decrease the offset from $\Delta_0$ to zero in 100$\,\mu$s. In a non-interacting system this experiment is a direct measurement of the tunneling of the array. By comparing our measurements to exact calculations of the tunneling dynamics, we conclude that the tweezers are balanced to within half a tunneling energy.

One other advantage of the tweezer array platform is the ability to easily realize arbitrary lattice geometries. In our one-dimensional system, we explore this by creating a lattice with staggered tunneling. To accomplish this, the radio-frequency tone difference between every other tweezer is changed to 4.2 MHz, with the edge pairs having a spacing of 4 MHz. This leads to a science configuration where the tunneling is modulated on alternate bonds and hence, qualitatively different behavior than the uniform tunneling array. We study tunneling dynamics in this staggered lattice by initializing atoms on the edge site (Fig.~\ref{fig:4}(b)). Again, there is good agreement with simulations, with two ballistically propagating tunneling trajectories forming before the atoms reach the opposite side of the chain. 

In conclusion, we have shown that optical tweezer arrays can be used to prepare many-body states of lattice fermions. The key advantages over optical lattices are increased flexibility in engineering Hubbard models on arbitrary geometry lattices and, with full post-selection on the atom number and spin, the possibility of reaching very low entropy states limited only by adiabaticity of the preparation. It is straightforward to extend the eight-tweezer system we have studied to arrays of over a hundred tweezers as has been demonstrated by the Rydberg atom array community~\cite{scholl2020programmable, ebadi2020quantum}, especially since power requirements for loading the tweezers from a degenerate Fermi gas are about three orders of magnitude lower. In one dimension, the staggered tunneling array we realized can be used to study topological states in the Su-Schrieffer-Heeger-Hubbard (SSHH) model \cite{PhysRevB.22.2099, RevModPhys.91.015005, barbiero2018quenched}. In two dimensions, exact diagonalization becomes intractable for system sizes of about twenty tweezers.  To scale the tweezer array to two dimensions, we plan to use two crossed acousto-optic deflectors, with the second dimension introduced by stroboscopically switching between chains created using the approach described in this work. The repetition rate for creating such a 2D array can be two orders of magnitude higher than lattice bandgaps, which avoids Floquet heating. This will allow studying 2D Hubbard models with arbitrary software-defined geometry. 

We would like to thank Lysander Christakis, Lawrence Cheuk, and Max Prichard for helpful discussions. This work was supported by the NSF (grant no. 2110475), the David and Lucile Packard Foundation (grant no. 2016-65128), and the ONR (grant no. N00014-21-1-2646).

\bibliography{bib}

\setcounter{figure}{0}
\setcounter{equation}{0}
\setcounter{section}{0}

\clearpage
\onecolumngrid
\vspace{\columnsep}

\newcolumntype{Y}{>{\centering\arraybackslash}X}
\newcolumntype{Z}{>{\raggedleft\arraybackslash}X}

\newlength{\figwidth}
\setlength{\figwidth}{0.45\textwidth}

\renewcommand{\thefigure}{S\arabic{figure}}
\renewcommand{\theHfigure}{Supplement.\thefigure}
\renewcommand{\theequation}{S\arabic{equation}}
\renewcommand{\thesection}{\arabic{section}}

\begin{center}
	\large{\textbf{Supplementary Information}}\\~\\
	
	\normalsize{Benjamin M. Spar, Elmer Guardado-Sanchez, Sungjae Chi, Zoe Z. Yan, Waseem S. Bakr\\
    \textit{Department of Physics, Princeton University, Princeton, New Jersey 08544, USA}
    
}
\end{center}
\section{Tweezer Loading}\label{sec:loading}

\begin{figure}[b]
\includegraphics{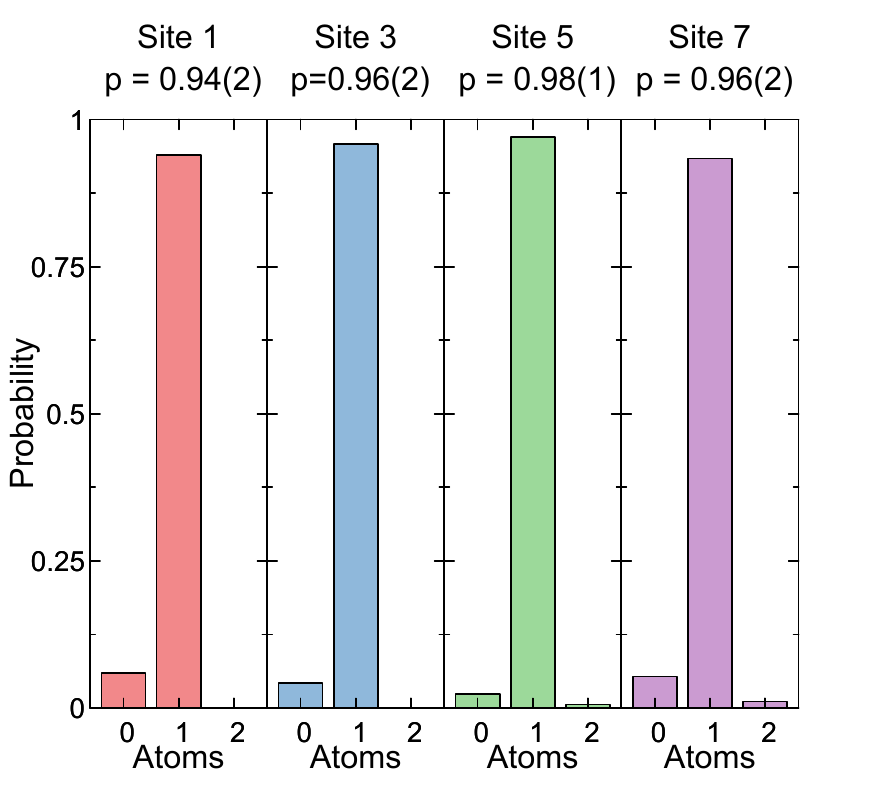}
\caption{\label{fig:4tweezerloading} Probability of detecting a spin up atom in each of the four loading tweezers for an experimental run of 167 shots.}
\end{figure}

 The experiment starts by loading the motional ground state of four independent tweezers with high fidelity. To characterize the loading fidelity, we resonantly remove state $|3\rangle$ atoms and image the state $|1\rangle$ occupation of the four sites (see Fig.~\ref{fig:4tweezerloading}, states are numbered starting from the lowest hyperfine state of the atom). For an experimental run of 167 shots, we image one atom on a given site with a fidelity of $96.0(8)\%$. Accounting for the measured fluorescence imaging fidelity, we have a single spin occupation of $97.5(9)\%$, corresponding to a two atom ground state loading fidelity of $95(2)\%$. This corresponds to an initial entropy of 0.12(3) $k_\mathrm{B}$ per particle, which we are able to reduce using post-selection. 

\section{Imaging}
\paragraph{Raman imaging:} For all data in the paper except in Fig.~2, we use the following imaging scheme. 
The atoms in the science configuration have their dynamics frozen by introducing a 5\% energy offset on every other site within 100~$\mu$s.
The tweezer depths are ramped in 1~ms to a higher value, typically three times the science depth.
The state $|3\rangle$ atoms are optically removed with a resonant light pulse.
To transfer the remaining state $|1\rangle$ atoms from the tweezers into the super-resolved optical lattice, the optical lattice is first slowly turned on to 20~$E_r$ in 5ms, then quickly ramped in 30~$\mu$s to 2500 $E_r$ to freeze dynamics.
The tweezers are turned off, and Raman imaging of state $|1\rangle$ atoms is performed~\cite{brown2017spin}.  The image fidelity is measured to be $98.5(4)$\%.
When two atoms end up in the same imaging lattice site before Raman imaging (either from imperfect loading of two $|1\rangle$ atoms one of which is in higher axial state of the tweezer or from imperfect removal of $|3\rangle$ atoms), light-assisted collisions from Raman imaging lead to parity-projected loss of the both atoms.  In rare instances (with probability $0.4(3)\%$), the two atom will register to adjacent sites which we can account for in the image reconstruction. 

\paragraph{Singles parity imaging:}
\noindent 
One characteristic of forming a Mott insulating state is the suppression of double and zero occupancies.  
To characterize the transition into the Mott regime, as discussed in Fig.~2 of the main text, we use an alternate parity imaging scheme that accurately detects the singles occupancy by transferring all doubles to a molecular state, which is deterministically lost to light-assisted collisions during the Raman imaging step.
First, after the ramp to the science tweezer configuration is finished, we transfer all atoms in state $|3\rangle$ to state $|2\rangle$ with a rf pulse at 594\,G.  
The magnetic field is ramped to 730\,G in 20 ms, where $a_{13}\,{=}\,-6900\,a_0$ and $a_{23}\,{=}\,2500\,a_0$.  
Here we have crossed the $|1\rangle-|3\rangle$ Feshbach resonance without exciting any higher vibrational states.
Next, we transfer all atoms in state $|2\rangle$ to state $|3\rangle$.
Finally, we ramp the field down across the resonance to the molecular side, associating all $|1,3\rangle$ doublons into dimer molecules.
Raman imaging is performed as described in the above section, but the dimers are always lost due to light-assisted collisions, leading to perfect parity imaging that can discriminate the number of singles.

%

\section{Tweezer Balancing Algorithm}
To balance the tweezer depths in the science configuration to less than $t$, we perform the following 2-step algorithm at $U/t \sim 1$. First, we balance the depths for the loading tweezers (which we label site 1, 3, 5, and 7) to within $\hbar\omega_z$ by ensuring our gradient spill procedure can produce equal loading probabilities on those four sites.  We measure the intensities on a camera (iXon Ultra 897) to obtain target intensities $(I_1,I_3,I_5,I_7)$ at the balanced value. Then, we repeat the process loading into tweezers 2, 4, 6, and 8, with the previous loading sites off. When we turn on all eight tweezers in the science configuration, we apply feedback on the eight control voltages to the AOM $V_1...V_8$ until the relative camera intensities match the recorded $I_1...I_8$ values.

\begin{figure}[tb]
\includegraphics[width = 4 in]{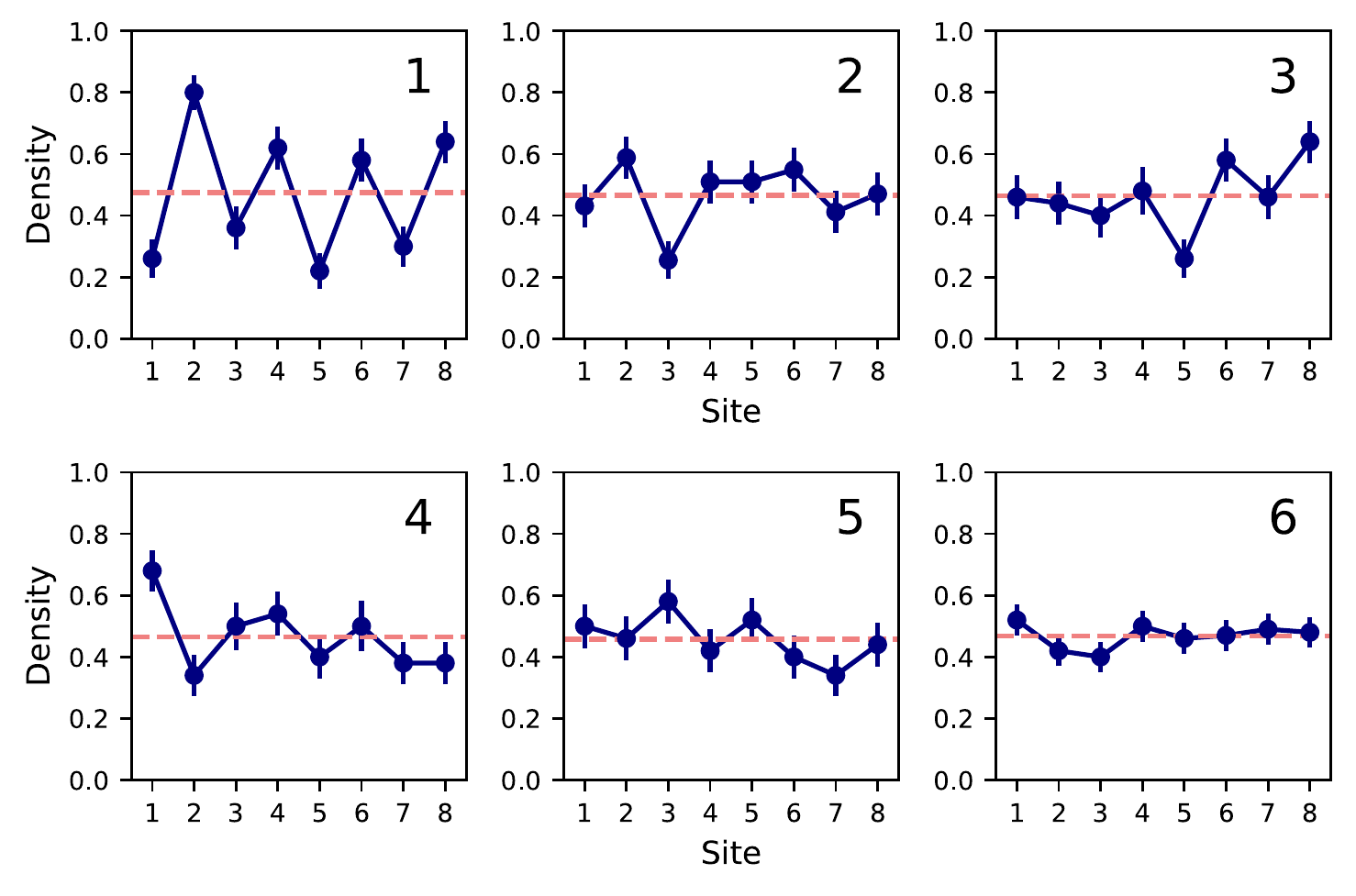}%
\caption{\label{fig:balancing} Evolution of density profiles in the science configuration during the tweezer offset balancing algorithm. Each panel shows the average occupation in the array in the science configuration at a different iteration of the algorithm, with feedback on the control voltages $V_i' = V_i - \alpha(\langle n_i \rangle-\langle n \rangle)$, where $\alpha$ is a constant proportionality factor.}
\end{figure}

This initial algorithm balances the depths to within $\hbar\omega_z$ but not to within $t$, a constraint an order of magnitude more stringent. Our final step is to have the atoms tunnel in the array and to minimize the disorder based on achieving a uniform density.  First loading in the odd sites, we adiabatically ramp on the even sites as described in the main text. At zero disorder we expect equal population among all sites. On every other shot, we load the even sites with the odd sites initially off, then adiabatically ramp on the odd sites, in order to avoid configurations in which certain sites can support tunneling into higher bands.  We take approximately 50 images, read out the site resolve density $\langle n_1 \rangle ... \langle n_8 \rangle$, and apply feedback on the control voltages proportional to $\langle n_i \rangle-\langle n\rangle$, where $ n=\frac{1}{8}\Sigma_i n_i$.  The process is repeated until the densities are balanced to within one standard deviation, as shown in Fig.~\ref{fig:balancing}. Convergence usually occurs within 10 cycles of this feedback loop and is limited primarily by radio-frequency cross-couplings between different tones generating the tweezers.

\section{Calculation of Hubbard Parameters}

To measure the on-site interaction $U$, we perform radio-frequency spectroscopy and resolve the energy shift between singly and doubly occupied tweezer sites~\cite{brown2017spin}.

We extract $t$ from the tunneling dynamics experiment shown in Fig.~4 of the main text.  The Python package QuSpin \cite{weinberg2019quspin} is used to simulate tunneling of a single atom across the array, assuming no disorder in the tweezer depths, and a least-squares fit to the data is performed with tunneling as the only free parameter. Errorbars are extracted from 500 bootstrapped samples of the experimental data.

\section{Entropy and Post-selection }\label{sec:entropy}
The two main sources of entropy in the correlated states prepared in the tweezer array are loading imperfections and entropy gained during the ramp to the correlated state.

The initial loading entropy is due to loading fewer than two atoms per site.  Entropy per particle before post-selection is given by
\begin{equation}
    \frac{S}{\langle N \rangle} = -\frac{k_\mathrm{B}}{2p}\left( p^2 \log \left(p^2\right) + 2p(1-p) \log \left(p(1-p)\right)+(1-p)^2\log \left((1-p)^2\right)\right)
\end{equation}
where $p$ is the single particle loading fraction after blowing one spin state. Although we have an entropy of 0.12(3) $k_\mathrm{B}$ after loading, due to the approximately 2 percent atom loss we experience over the course of our experiment, the effective loading entropy before post-selection is 0.19(4) $k_\mathrm{B}$ per particle.

The second source of entropy comes from the ramp that turns on the non-loading sites, ending in the balanced 8-site science configuration. Our ramp parameters are chosen to be as adiabatic as possible subject to lifetime constraints, with over 98\% state overlap calculated from the perfect 4-site initial state to the 8-site ground state at $U/t=6.7(3)$. This procedure should only add $0.04\,k_\mathrm{B}$ per particle of entropy for the perfect loading state of two atoms per site. However, when we have an imperfect loading state (i.e. at least one loading site has less than two atoms), the ramping procedure is no longer isentropic. 

The problem of imperfect loading is partially mitigated by post-selection of perfect occupation of one spin state, \textit{i.e.} imaging four spin-up particles. The entropy is then only from uncertainty in the spin-down population, such that
\begin{equation}
    \frac{S_{\mathrm{post-selected}}}{\langle N\rangle} = -\frac{k_\mathrm{B}}{1+p}\big( p\log p+(1-p) \log (1-p) \big).
\end{equation}
With post-selection the loading entropy per particle (accounting for atom loss) decreases to  $0.09(1)\,k_\mathrm{B}$, and the lower bound for the final entropy after the ramp is $0.18(2)\,k_\mathrm{B}$ per particle.  With future implementation of full spin-charge readout, post-selecting on perfect loading of both spin states will decrease the loading entropy to zero and the ramping entropy to $0.04\,k_\mathrm{B}$.

To determine the final entropy for correlated states as in Fig.~3 of the main text, we subject the antiferromagnetic correlations to a two-parameter fit of temperature and chemical potential, assuming the many-body statistics are described by the grand-canonical ensemble.
The grand canonical partition function sums over the microstates
\begin{equation}
    Z(\mu,T) = \sum_\Omega\exp\left(\frac{ \mu N_{\Omega} - E_{\Omega}}{k_\mathrm{B}T}\right)
\end{equation}
where each microstate is labelled by $\Omega$ and has total particle number $N_{\Omega} = 4 + N_{\Omega,|\downarrow\rangle}$ and energy $E_{\Omega}$.
We fit the spin-up correlations to those computed by numerics for a wide range of $(\mu,T)$, subject to the constraint that the measured atom number $\langle N \rangle$ with post-selection must equal the grand canonical expectation of the atom number $\langle N_{\rm GC} \rangle$,
\begin{align}
    \langle N \rangle &= 4 + 4 p \\
    \langle N_{\rm GC} \rangle &=\frac{1}{Z(\mu,T)}\sum_{\Omega}N_\Omega e^{-(E_\Omega - \mu N_\Omega)/k_\mathrm{B}T}
\end{align}

We do a weighted least squares fit on all correlators $C_{ij}$ between tweezer sites $i$ and $j$, where is the weighting is done by the variance of the covariance of $C_{ij}$. 
With the best-fit values of $\mu'$ and $T'$, entropy is then
\begin{align}
    S_{\rm GC,\mathrm{post-selected}} &=-k_\mathrm{B}\sum_{\Omega} P_{\Omega,\rm GC}\log P_{\Omega,\rm GC} \\
    P_{\Omega,\rm GC}(\mu',T')& = \frac{1}{Z(\mu',T')}e^{-(E_{\Omega} - \mu' N_{\Omega})/k_\mathrm{B}T'}
\end{align}

\section{Bootstrapping Temperature }\label{sec:boottemp}

\begin{figure}[tb]

\includegraphics[width = 4 in]{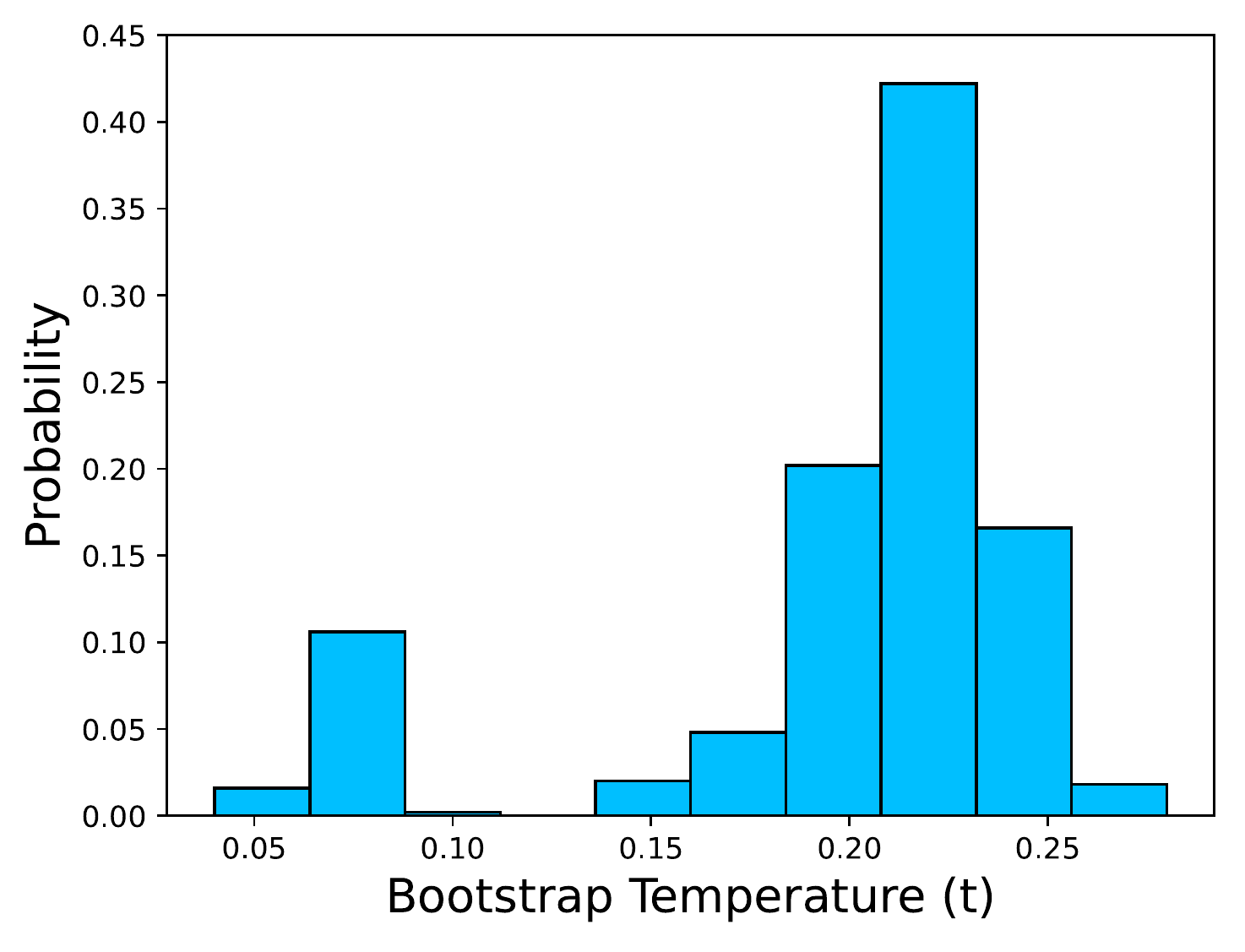}
\caption{\label{fig:boothist} Histogram of the temperature fit for 500 bootstrapped samples. There are two distinct peak instead of one, suggesting that the fit fails for low temperature samples. At the lower entropies that can be obtained by post-selecting on both spin states, new methods will need to be developed to measure temperature.}
\end{figure}

For the 8-site one dimensional Fermi-Hubbard model, the spin correlators $C_{ij}$ show strong finite size effects and are not all monotonic with temperature. Additionally, at low temperatures, the correlators begin to saturate with decreasing temperature. As such, for low enough temperatures, spin correlations are no longer a good thermometer for the system. At the temperatures reached in the experiment, we begin to see failure of this fitting method by bootstrapping a dataset with 601 runs (Fig. ~\ref{fig:boothist}). The temperature fits have a bimodal distribution, and around ten percent of bootstrapped samples have temperatures lower than the temperature corresponding to just the loading entropy ($k_\mathrm{B} T = 0.08(1) k_\mathrm{B}$). This suggests that for the colder temperature bootstrapped samples, the fit is failing. However, these failures are only a small fraction of samples.

\clearpage
\end{document}